\documentclass[letterpaper]{article}

\usepackage[T1]{fontenc}

\usepackage{geometry}
\usepackage{setspace}

\usepackage[style=chem-acs, articletitle=true]{biblatex}
\usepackage{graphicx}
\usepackage{float}
\newfloat{scheme}{htbp}{los}
\floatname{scheme}{Scheme}
\floatname{chart}{Chart}
\newfloat{graph}{htbp}{loh}

\usepackage{chemformula} % Formulas using \ch{}
\usepackage[version = 4]{mhchem} % Formulas using \ce{}

\usepackage{amsmath, amssymb}
\usepackage{cases}
\usepackage{braket}
\usepackage{siunitx}

\usepackage{hyperref}
\hypersetup{
  pdfpagemode=, 
  pdfstartpage=1,
  pdfstartview=FitH,
  pdfmenubar=true,
  pdftoolbar=true,
  colorlinks=true,
  linkcolor=blue,
  citecolor=blue,
  bookmarksopen=false,
}

\renewcommand{\vec}{\boldsymbol}

\usepackage{xcolor}

\usepackage{authblk}
\author[1]{Alexander Blech}
\author[1]{Monika Leibscher}
\author[1]{Christiane P. Koch*}
\affil[1]{
Freie Universit\"at Berlin, Dahlem Center for Complex Quantum Systems \& Fachbereich Physik, Arnimallee 14, D-14195 Berlin, Germany
}

\title{
Chiral rotational dynamics in the molecular frame: Breaking symmetry with angular momentum
}
\date{*Email: christiane.koch@fu-berlin.de}

\begin{document}

\maketitle

\begin{abstract}
Achiral molecules can be prepared in chiral superposition states, with handedness carried by molecular rotation rather than structure. We derive the conditions for such chiral rotational dynamics from the requirement to break the continuous spatial rotational symmetry and the molecular point group symmetry.
In an ensemble of randomly oriented molecules, net handedness requires excitation with three non-coplanar polarization directions, realized by microwave, THz, or combined optical and THz pulses. Because the chirality resides in the molecular-frame distribution of angular momentum, no laboratory-frame alignment or orientation is needed.
The induced chirality comprises a time-odd and a time-even contribution, which a phase-sensitive microwave probe detects separately. Photoelectron circular dichroism senses only the time-even contribution but accesses the persistent time-odd one after a delayed non-resonant pulse.
Our results show how to distinguish time-odd from time-even chirality and provide a source of time-odd chirality for driving enantio-selectivity far from equilibrium, without truly chiral reagents.
\end{abstract}

Molecular chirality, or handedness, is defined by the absence of any form of inversion or reflection symmetry.
Beyond this purely geometrical perspective, a further distinction arises from behavior under time reversal:
\emph{True} (time-even) chirality refers to chiral structures or states whose handedness remains invariant under time reversal, whereas \emph{false} (time-odd) chirality switches handedness \cite{Barron1986}.
A parity-violating energy difference due to the weak interaction is predicted only for truly chiral systems \cite{Barron1986a,Quack2022} and enantio-selectivity in thermal equilibrium requires truly chiral reagents \cite{Barron1987}.
Far from equilibrium, however, time-odd chirality can induce transient, dynamic enantio-selective effects \cite{Barron2013,BanerjeeGhosh2018} and has been utilized to drive absolute asymmetric synthesis, i.e., generate enantiomeric excess in a chemical reaction that would otherwise yield a racemic mixture \cite{Micali2012}.
Yet, in the growing toolbox for controlling~\cite{Lao2025,Li2026,Raucci2022,Liu2025,Ayuso2022} and creating~\cite{Tanioka2026,Xu2020other,Zhu2018,Romao2024,Du2026,Tikhonov2022,Leibscher2024,Mayer2022,Chen2024,Moitra2025,Ishitobi2026} handedness, time-odd chirality offers a resource that remains to be fully explored.

Dynamical chirality has become an active field in solid-state physics, with realizations in chiral electronic phases in light-driven achiral crystals~\cite{Xu2020other},
optomechanical systems with tunable optical chirality~\cite{Du2026} and chiral phonons~\cite{Zhu2018,Romao2024,Juraschek2025}.
For phonons, circular atomic motion generates handedness when combined with a polar direction perpendicular to its plane: a time-even direction, such as a polar crystal axis, yields false chirality~\cite{Gill2026}, whereas a time-odd direction, such as the linear momentum of helical phonons, yields true chirality~\cite{Ishito2023}.
In isolated molecules, which offer a versatile platform for exploring fundamental symmetry and chiroptical phenomena, experimental realization of false chirality has remained elusive.
While photoionization has been shown to access handedness beyond that encoded in the equilibrium structure~\cite{Tsitsonis2026}, it remains an open question which spectroscopic techniques are sensitive to time-odd chirality, ideally discriminating it from its time-even counterpart.

In gas-phase samples of atoms or molecules, theoretical proposals for laser-induced chirality have primarily focused on
vibrational~\cite{Tikhonov2022,Leibscher2024} and electronic~\cite{Mayer2022,Chen2024,Moitra2025,Ishitobi2026} dynamics.
Rotational motion provides a complementary route to molecular handedness.
Rotational enantiomers are commonly defined by opposite senses of rotation~\cite{Atkins1976} and are thus falsely chiral since angular momentum is odd under time reversal.
Yet the rotation axis itself is invariant under time reversal, so that a rotation about axes tilted out of a mirror plane is truly chiral, as we show below.
Rotational coherences are long-lived, persisting well beyond typical electronic and vibrational timescales.
At high angular momentum, rotational levels can form bands of near-degenerate states of opposite parity~\cite{Harter1984}, enabling the formation of chiral superpositions \cite{Bunker2004}.
Such chiral cluster states can be prepared by combining an optical centrifuge with a static electric field~\cite{OwensPRL2018}.
There, however, the angular momentum is locked to the propagation direction of the field, tying the handedness to a directional bias in the laboratory frame.
At low angular momentum, in contrast, the handedness of a rotational wave packet can be identified with that of the three-dimensional trajectory traced out by the expectation value of a molecular axis~\cite{Lohr2026}, characterized by two pseudoscalars, a dynamical one that is odd under time reversal and a geometric one that is even. Since this requires a time-dependent orientation of the molecular axis, the handedness is, again, tied to the laboratory frame.

Here, we derive the general symmetry conditions under which rotational motion renders an achiral molecule chiral.
The chirality resides in the distribution of the angular momentum in the molecular frame, with no preferred laboratory direction, and comprises two contributions: a time-odd one, the orientation of the angular momentum along a molecular axis, and a time-even one, the tilt of the angular momentum distribution out of the molecular mirror plane.
We show that generating net handedness in an ensemble of randomly oriented nonlinear rotors requires interaction with electric fields along three non-coplanar polarization directions, with the accessible classes of molecules determined by the number of non-vanishing dipole components.
Further, we present two possible realizations involving only low-lying rotational levels --- resonant microwave excitation, which generates a coherent wave packet whose chirality alternates between time-odd and time-even character, and THz or combined non-resonant optical and THz excitation, after which only the time-odd contribution persists.
A phase-sensitive microwave probe, adapted from three-wave mixing of chiral molecules~\cite{Patterson2013,Eibenberger2017,Perez2017}, resolves the two contributions separately, whereas photoelectron circular dichroism (PECD) \cite{Sparling2025} is sensitive to the time-even one, as confirmed by our time-dependent simulations.
Preparing and discriminating the two forms of chirality in the molecular frame opens a practical route to investigating time-odd chiroptical phenomena in gas-phase samples of randomly oriented molecules.

\begin{figure}
\centering
\includegraphics[scale=1]{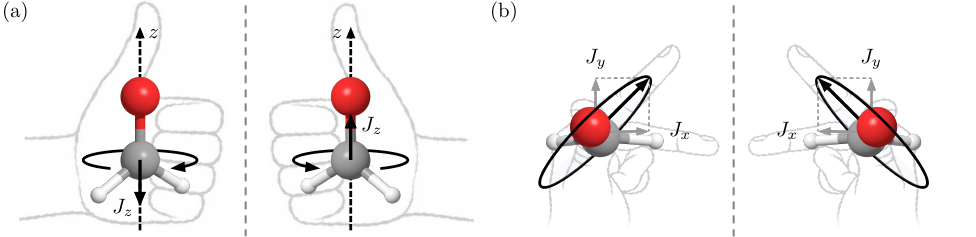}
\caption{
A rotating achiral molecule, here \ce{CH2O}, becomes dynamically chiral.
\textbf{(a)}
When angular momentum is localized along the molecule's $z$-axis, the sense of rotation reverses upon reflection, breaking the molecule's mirror symmetry. Handedness can be assigned via the convention of the thumb oriented along the molecular symmetry axis (dashed arrow), and the fingers represent the sense of rotation encoded in the angular momentum component $J_z$ (solid arrow).
Time reversal also reverses the sense of rotation, so this chirality is time-odd.
\textbf{(b)}
When angular momentum is distributed along all directions within a plane (black circle) that is tilted with respect to the molecule's symmetry plane, 
its mirror image has the opposite tilt and cannot be rotated back onto the original; the two tilts thus form an enantiomeric pair.
One representative direction (black arrow) is shown together with its components $J_x$ and $J_y$ (grey arrows).  Its handedness follows from a tripod of three fingers: two lie in the molecular plane and one is along the tilt of the angular-momentum plane. Time reversal flips the sign of both $J_x$ and $J_y$, which leaves the tilt unchanged, so this chirality is time-even.
}
\label{fig:sketch}
\end{figure}

\paragraph*{Conditions for creating rotational chirality}
Handedness can arise either from three non-coplanar polar vectors or from a polar vector and a collinear pseudovector \cite{Barron2020}.
Figure~\ref{fig:sketch} illustrates this for the combination of a molecular axis and the molecule's rotational angular momentum.
In Fig.~\ref{fig:sketch}(a), an achiral molecule becomes dynamically chiral when rotating about an axis parallel to a mirror plane, or their intersection if multiple such planes exist.
We define this axis as the molecule-fixed quantization axis, $z$.
The corresponding chiral-sensitive pseudoscalar is the angular momentum projection onto the molecular $z$-axis, $\hat J_z$.
It switches sign under reflection at the mirror planes, breaking the molecule's reflection symmetry and --- in combination with the nuclear arrangement --- producing a chiral molecular state.
A necessary requirement is that the nuclear scaffold defines an oriented molecular axis%
\footnote{
If the axis can only be aligned, reversing the molecular axis would compensate the sign change, rendering the rotational state equivalent to its mirror image.}
such that rotational enantiomers can be created in molecules without a center of inversion.
Time-reversal changes the sense of molecular rotation, interchanging the enantiomers and reflecting the dynamical, time-odd nature of this form of rotational chirality.
The symmetrized product $\{\hat J_x,\hat J_y\}=\hat J_x\hat J_y+\hat J_y\hat J_x$ is likewise odd under reflection at the mirror plane but, being quadratic in the angular momentum, even under time reversal, i.e., truly chiral. Its expectation value signals an angular momentum distribution that is tilted out of the mirror plane, with a definite sense of the tilt rather than a definite sense of rotation, cf. Fig.~\ref{fig:sketch}(b). Both pseudoscalars are molecular-frame properties, independent of any laboratory direction.
We quantify the induced chirality by the complex pseudoscalar
\begin{equation}
\chi=
i\braket{\hat J_z}/\hbar-\braket{\{\hat J_x,\hat J_y\}}/\hbar^2
\,,
\label{eq:chi}
\end{equation}
whose imaginary and real parts are the time-odd and the time-even contribution and whose modulus measures the total degree of dynamical chirality.
In general, the chirality of a rotational state is contained in the part of its density matrix that is odd under reflection at the molecular mirror plane, which may also comprise higher-order correlations of the angular momentum components. Equation~\eqref{eq:chi} retains the two lowest orders, linear and quadratic in the angular momentum.
These are dynamically linked by the Heisenberg equation, which for $\hat J_z$ is Euler's equation for the rigid rotor,
\begin{equation}
\frac{d}{dt}\braket{\hat J_z}=-\frac{B_x-B_y}{\hbar^2}\braket{\{\hat J_x,\hat J_y\}}+\braket{\hat N_z}\,,
\label{eq:euler}
\end{equation}
with $B_x$, $B_y$ the rotational constants about $x$ and $y$ and $\hat N_z$ the torque exerted by external fields about $z$. The time-odd contribution is thus generated either by an external torque about the chiral axis or, for an asymmetric rotor, from the time-even contribution.

\begin{figure*}
\centering
\includegraphics[width=\textwidth]{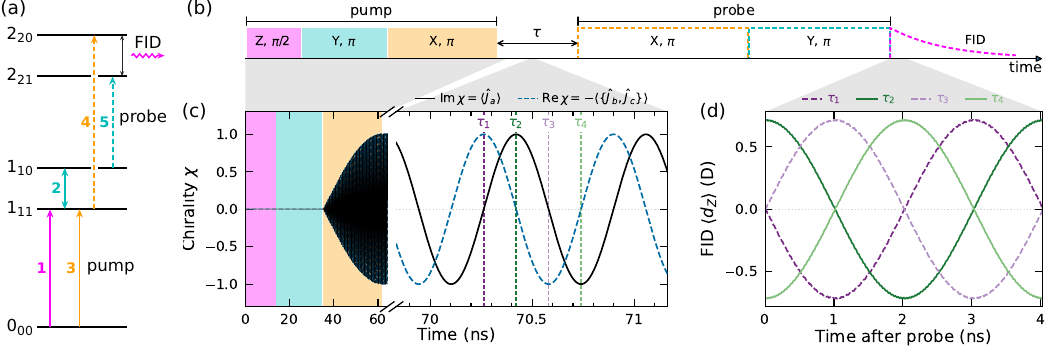}
\caption{
Rotational chirality induced in an achiral asymmetric rotor, here \ce{COFCl}, by a sequence of three microwave pulses, polarized along three mutually orthogonal directions, and its detection by a microwave probe.
\textbf{(a)} Rotational levels and excitation scheme with initial state $0_{00}$, pump transitions 1--3 (solid arrows), probe transitions 4 and 5 (dashed arrows), and detected free induction decay (FID, wavy arrow); colors indicate the polarization direction.
\textbf{(b)} Pulse sequence with polarization directions, pulse areas, and pump-probe delay $\tau$.
\textbf{(c)} Time-even and time-odd contributions to the induced chirality, $\mathrm{Re}\,\chi$ and $\mathrm{Im}\,\chi$, cf. Eq.~\eqref{eq:chi}, during the pump pulses (shaded) and, on an enlarged time scale, during the subsequent field-free evolution. Four delays are marked by $\tau_1$ to $\tau_4$ corresponding to purely time-even and time-odd chirality for the two handednesses.
\textbf{(d)} Simulated FID along $Z$ after the probe pulses for the four pump-probe delays marked in (c) (decay by collisions and Doppler dephasing not included).
Its phase reveals both the handedness and the time-even or time-odd character of the chirality.
}
\label{fig:resonant}
\end{figure*}

To derive the conditions for creating non-zero $\braket{\hat J_z}$, we denote
the quantum numbers labeling the projection of the angular momentum onto the molecule-fixed and laboratory-fixed quantization axis by $K$ and $M$, as usual.
With our choice of axes, states with $+K$ and $-K$ correspond to the two enantiomeric senses of rotation.
For symmetric tops, $K$ is a good quantum number and states with the same $|K|$ are degenerate. As a result, any thermal state is an equal mixture of $+K$ and $-K$ states, i.e., racemic.
In asymmetric top molecules, $K$ is not a good quantum number, but since the eigenstates are symmetric or antisymmetric superpositions of $K$ and $-K$ states \cite{Bunker2012}, they likewise constitute a racemic mixture.
Preparing a state with net chirality requires 
(i) creating a coherent superposition that breaks the $\pm K$ symmetry,
and (ii) preventing the cancellation of contributions to $\hat J_z$ from states with different $M$.
Since $\{\hat J_x,\hat J_y\}$ transforms like $\hat J_z$ (cf. Supporting Information), the conditions derived below apply to both.

Condition (i) is fulfilled for superpositions consisting of rotational eigenstates $\ket{j}$ and $\ket{j'}$ with $\braket{j|\hat J_z|j'}\neq0$.
Since $\hat J_z$ and the corresponding dipole moment component transform identically under the molecular rotation group (cf. Supporting Information), the symmetry requirements for such superpositions are the same as those for non-vanishing transition dipole moments $\braket{j|\hat d_z|j'}$.
The number of non-zero dipole moment components determines the rotational dipole selection rules and can thus be used to create a chiral rotational distribution in a given molecule. 
For molecules with at least two non-vanishing dipole components, chiral rotational superpositions can be prepared from arbitrary initial states, provided a sufficient number of pulses and polarization directions are employed \cite{Pozzoli2022}.
For molecules with only a single non-zero dipole component, this is possible only for specific initial states.
An exception are symmetric top molecules.
Since their dipole moment is parallel to the molecule's symmetry axis, rotational transitions are only allowed if $\Delta K=0$ \cite{Gordy1984}.
Breaking the $\pm K$ symmetry in symmetric tops therefore requires transition dipole moments perpendicular to the permanent dipole moment and thus excitation via vibrational or electronic states.

Condition (ii) encodes the creation of macroscopic enantiomeric excess and thus corresponds to addressing the spatial degeneracy of rotational states.
Since the direction of the induced rotation depends on molecular orientation, averaging over the random orientations of a gas phase ensemble (i.e., summing over $M$) does not result in a net chirality unless the excitation sequence also breaks space-inversion symmetry in the laboratory frame.
We show in the Supporting Information that this is equivalent to at least three pulses with non-coplanar polarization directions.
Related requirements arise for chiral rovibrational wave packets \cite{Leibscher2024}, synthetic chiral light \cite{Ayuso2019,Neufeld2020,Ayuso2021}, and enantio-selective population transfer \cite{Leibscher2019,Eibenberger2017,Perez2017}.
With microwave pulses, the three fields must be phase-locked and resonant with individual rotational transitions, as in enantio-selective population transfer, but, unlike synthetic chiral light, require no fixed frequency ratios. In contrast to population transfer, which enriches a chiral molecule's rotational state using three dipole components in a closed cycle, inducing rotational chirality needs just one.

\paragraph*{Creating and detecting rotational chirality with microwave fields}
As a concrete example of an achiral asymmetric top molecule, we consider carbonyl chloride fluoride (\ce{COFCl}), which has $C_{\mathrm{s}}$ point symmetry and thus two non-vanishing permanent dipole components.
Associating the molecule-fixed $z$-axis with the $a$-principal axis of inertia (corresponding to convention I from Refs.~\cite{Bunker2012,Zare1988}), a chiral rotational wave packet can be constructed from states connected by $a$-type transitions.
When labeled in terms of the spectroscopic notation, $\ket{j}=\ket{J_{K_a,K_c}, M}$, this corresponds to states with $\Delta K_a$ even and $\Delta K_c$ odd  \cite{Gordy1984,Bunker2012}, and the simplest realization is a superposition of $1_{10}$ and $1_{11}$.
Starting from the rotational ground state, $0_{00}$, this wave packet can be created with the excitation scheme shown in Fig.~\ref{fig:resonant}(a,b).
It consists of a $Z$-polarized $\pi/2$-pulse exciting half of the ground state population to state $1_{11}$, a $Y$-polarized $\pi$-pulse transferring the complete population from state $1_{11}$ to $1_{10}$, and an $X$-polarized $\pi$-pulse driving the remaining ground state population to $1_{11}$.
The corresponding dynamics is simulated by solving the time-dependent Schrödinger equation for a rigid rotor in the electric dipole approximation (see Supporting Information for computational details).
After the third pulse, the angular momentum periodically localizes along the molecular $a$-axis, and the resulting time-odd handedness, expressed in the sign of $\braket{\hat J_a}$, oscillates in time, cf. Fig.~\ref{fig:resonant}(c).
The time-even contribution, $-\braket{\{\hat J_b,\hat J_c\}}$, oscillates with the same amplitude, a quarter period ahead as dictated by Eq.~\eqref{eq:euler}.
Both contributions reside in the coherence between $1_{11}$ and $1_{10}$ and Eq.~\eqref{eq:chi} becomes $\chi(t)=e^{i(\omega_\chi t+\varphi)}$, with $\hbar\omega_\chi$ the splitting of the two levels.
Thus, $|\chi|$ is constant and only the character of the chirality alternates between a definite sense of rotation and of tilt, cf. Fig.~\ref{fig:sketch} and Fig.~S1 of the Supporting Information.
The phase of this oscillation --- and thus the chirality at a given time --- can be controlled by the relative phase of the microwave pulses.
A shift of $\pi$ reverses the sign of both $\braket{\hat J_a}$ and $\braket{\{\hat J_b,\hat J_c\}}$, i.e., the handedness, while a shift of $\pi/2$ interchanges time-odd and time-even chirality.
Handedness can also be reversed by an odd permutation of the pulse polarization directions.

Since both pseudoscalars are molecular-frame properties, detecting them in a randomly oriented ensemble requires a probe that itself breaks space-inversion symmetry in the laboratory frame.
For the coherent wave packet, this can be achieved by measuring the free induction decay (FID) following a microwave probe adapted from three- and six-wave mixing of chiral molecules~\cite{Patterson2013,Eibenberger2017,Perez2017,Sun2023}.
The superposition of $1_{11}$ and $1_{10}$, which carries $\chi$, does not radiate, since contributions from states with opposite $M$ cancel.
After a delay $\tau$, two resonant $\pi$ pulses with polarization directions $\vec e_1$ and $\vec e_2$ therefore transfer the superposition to another pair of levels, here $2_{20}$ and $2_{21}$, cf. Fig.~\ref{fig:resonant}(a,b). This converts the chirality into a macroscopic polarization, which gives rise to an FID along $\vec e_3$,
\begin{equation}
\braket{\hat d_{3}}(t)
= d\,\left[\vec e_1\cdot(\vec e_2\times\vec e_3)\right]\,\mathrm{Re}\big[\chi(\tau)\,e^{-i\theta(t)}\big]
= d\,\left[\vec e_1\cdot(\vec e_2\times\vec e_3)\right]
\Big[
\frac{\braket{\hat J_z}_\tau}{\hbar}\sin\theta(t)
-
\frac{\braket{\{\hat J_x,\hat J_y\}}_\tau}{\hbar^2}\cos\theta(t)
\Big]
\,.
\label{eq:fid}
\end{equation}
Here, $\chi(\tau)$ and the expectation values $\braket{\cdot}_\tau$ refer to the wave packet at the arrival of the probe.
The triple product $\vec e_1\cdot(\vec e_2\times\vec e_3)$ reflects the symmetry breaking required of the probe and vanishes if the two probe polarizations and the detection direction are coplanar.
The phase $\theta(t)=\omega_\mathrm{3}(t-\tau)+\theta_0$ evolves with the frequency $\omega_3$ of the transition between $2_{21}$ and $2_{20}$, not with $\omega_\chi$. $\theta_0$ is fixed by the phases of the two probe pulses, and $d$ is the matrix element of the dipole moment along $\vec e_3$ between the two states.
In Fig.~\ref{fig:resonant}, the probe pulses are polarized along $X$ and $Y$ and the FID is detected along $Z$, such that $\vec e_1\cdot(\vec e_2\times\vec e_3)=1$ and
$d=|d_a|\cos\vartheta/\sqrt3$, where $d_a$ is the permanent dipole moment along $a$ and $\tan2\vartheta=\sqrt3(B-C)/(2A-B-C)$.
The chirality at the delay $\tau$ is thus encoded in the phase of the FID:
Since $\chi(\tau)=e^{i(\omega_\chi\tau+\varphi)}$, the delay selects which contribution is probed, with the time-even contribution appearing as the $\cos\theta$ and the time-odd one as the $\sin\theta$ component, cf. Fig.~\ref{fig:resonant}(d), and reversing the handedness shifts the FID by $\pi$.
Separating the two contributions requires pump, probe and detection to be phase-locked~\cite{Patterson2013}. Without phase-locking between pump and probe, measuring the FID intensity still yields $|\chi|$.
Equation~\eqref{eq:fid} holds for probe polarizations with $\vec e_3$ perpendicular to the polarization of the last pump pulse. Otherwise, the FID contains additional contributions that stem from the orientation of the ensemble and do not vanish for achiral rotational states.
In contrast to three-wave mixing of chiral molecules, which also relies on three non-coplanar polarization directions, pump and probe here involve only $a$-type and $b$-type transitions.

\paragraph*{Creating rotational chirality with optical and THz fields and detecting it with PECD}

\begin{figure}[tbp]
\centering
\includegraphics[scale=1]{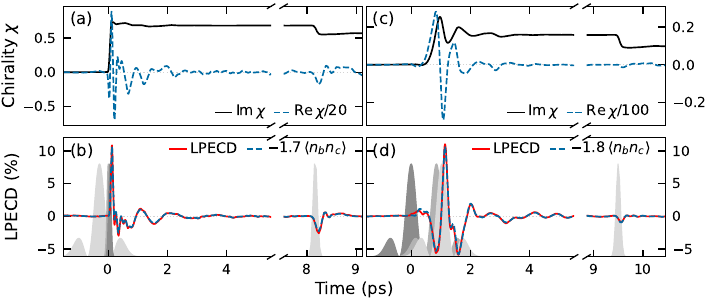}
\caption{
Rotational chirality induced in a thermal ensemble of achiral asymmetric rotors, here \ce{CH2O} at a temperature of \SI{5}{K}, by THz and optical pulses (gray shadings) with mutually orthogonal polarization directions:
\textbf{(a,b)} combination of a THz pulse with two femtosecond optical pulses;
\textbf{(c,d)} three THz pulses.
\textbf{(a,c)} Time-odd and time-even contributions to the induced chirality, $\mathrm{Im}\,\chi=\braket{\hat J_a}$ and $\mathrm{Re}\,\chi=-\braket{\{\hat J_b,\hat J_c\}}$ (scaled as indicated), cf. Eq.~\eqref{eq:chi}, with the molecular $a$-axis along the \ce{C-O} bond and $c$ normal to the molecular plane.
\textbf{(b,d)} Linear PECD (LPECD) assuming instantaneous ionization together with the cross-alignment $\braket{n_bn_c}$, scaled as indicated.
Behind the axis break, a delayed $X$-polarized optical pulse maps the persistent $\braket{\hat J_a}$ onto the LPECD.
}
\label{fig:nonresonant_pecd}
\end{figure}

Molecular-frame angular momentum can also be induced by combining non-resonant optical or infrared pulses and THz radiation.
Such excitation does not require individually resolved rotational transitions, and since the optical interaction is averaged over the optical cycle, neither optical carrier-phase locking nor phase locking between the optical and THz pulses is required, only control of the pulse delays, polarization directions and THz phase.
This scheme offers a qualitatively different form of rotational chirality compared to resonant microwave excitation. Whereas the latter produces a coherent wave packet with an oscillating handedness, the non-resonant pulse sequence generates a time-odd handedness that remains constant after the fields are switched off.
Figure~\ref{fig:nonresonant_pecd}(a) illustrates this for a thermal ensemble of formaldehyde (\ce{CH2O}) at \SI{5}{K},
using two delayed optical femtosecond pulses, polarized along the laboratory $X$- and $Y$-direction, combined with a THz pulse, polarized along the $Z$-direction.
In contrast to \ce{COFCl}, \ce{CH2O} has only a single permanent dipole-moment component, oriented along the \ce{C-O} bond, and therefore is not fully controllable.
The dynamics is simulated classically, which is justified for the timescales considered here, well before the quantum revivals, cf. Supporting Information.
The two optical pulses generate unidirectional rotation about the $Z$-axis (with $\braket{J_Z}\approx\SI{4}{\hbar}$) and a small degree of alignment in the $XY$-plane (less than \SI{2}{\%}).
This is reminiscent of twisted polarization pulses which also generate unidirectional rotation and, for chiral molecules, enantio-selective orientation~\cite{Yachmenev2016,Gershnabel2018,Milner2019}.
Since achiral molecules lack the required components in their polarizability tensor, optical, far off-resonant excitation alone is not sufficient to induce chirality (cf. Supporting Information). When combined with a THz pulse, which couples to the electric dipole moment, the sequence generates also a nonzero angular-momentum projection along the molecular axis (the \ce{C-O} bond), creating a chiral rotational state.
Here, $\braket{\hat J_a}$ stems from the torque in Eq.~\eqref{eq:euler}, exerted by the optical pulses through the anisotropic polarizability.
Figure~\ref{fig:nonresonant_pecd}(c) shows that excitation with three THz pulses with mutually orthogonal polarization directions also generates a rotational distribution with persistent handedness.
With the dipole of \ce{CH2O} parallel to $a$, the THz fields exert no torque about this axis. $\braket{\hat J_a}$ builds up entirely from the time-even contribution, the first term of Eq.~\eqref{eq:euler}, as the rotor asymmetry turns the field-induced tilt into a persistent sense of rotation.
In both cases, the time-even contribution dephases within a few picoseconds, since the wave packet comprises many rotational coherences with different frequencies, whereas $\braket{\hat J_a}$ persists, cf. Fig.~\ref{fig:nonresonant_pecd}(a,c). Hence, rotational dephasing does not erase the induced handedness but reduces it to its time-odd contribution, which, like the orientation induced by THz pulses~\cite{Xu2020,Tutunnikov2021}, persists until collisional relaxation.

In the broad rotational wave packet created by optical or THz pulses, chirality arises from the population of many levels. A microwave probe, however, is resonant with only a single transition and therefore would capture only a small fraction of it. PECD, in contrast, can monitor time-dependent chirality \cite{Beaulieu2017,Waters2022,Han2025} without being restricted to individual transitions.
Adapting the pump-probe scheme of Ref.~\cite{Tikhonov2022}, we propose to probe the rotational dynamics with a delayed femtosecond extreme-ultraviolet pulse, circularly polarized in the $XY$-plane.
Since the rotational state remains effectively frozen during the pulse, the ionization amplitude at time $t$ is
given by integrating over the orientation-dependent ionization amplitudes $a_{\vec k}(t,\omega)$, 
\begin{equation}
a_{\vec k}(t)=\int a_{\vec k}(t,\omega)
P_{\mathrm{rot}}(t,\omega)
\mathrm{d}\omega \;,
\label{eq:ionization_amplitude}
\end{equation}
where $\vec k$ denotes the photoelectron momentum in the laboratory frame, $\omega$ the Euler angles specifying the molecular orientation and $P_{\mathrm{rot}}=|\psi(t,\omega)|^2$ the rotational probability distribution.
Since $P_\mathrm{rot}$ is even under time reversal, PECD cannot probe $\braket{\hat J_z}$ directly. It arises instead from the (time-even) cross-alignment $\braket{n_x n_y}$, with $n_\alpha$ the component of the probe axis $Z$ along the molecular axis $\alpha$, which breaks the reflection symmetry of the orientation distribution with respect to the polarization plane. Rotation generates it according to
\begin{equation}
\frac{d}{dt}\braket{n_xn_y}
=\big\langle\Omega_z\,(n_y^2-n_x^2)\big\rangle+\big\langle n_z\,(\Omega_xn_x-\Omega_yn_y)\big\rangle\,,
\label{eq:kinematic}
\end{equation}
with $\Omega_\alpha$ the molecular-frame angular velocity, so PECD is indirectly sensitive to the time-odd dynamics---analogous to photoexcitation circular dichroism, where a chiral electronic current creates a (time-even) chiral charge distribution~\cite{Beaulieu2018}.

Figure~\ref{fig:nonresonant_pecd}(b,d) shows the linear PECD (LPECD), which quantifies the asymmetry of photoelectron emission into the forward and backward hemispheres with respect to the polarization plane of the ionizing pulse \cite{Lux2015}.
It follows the cross-alignment $\braket{n_bn_c}$ and emerges during the creation of the net angular momentum orientation, reaching a maximum strength of about \SI{11}{\%}, which is clearly within the capability of photoelectron detection techniques \cite{Kastner2016}, and decays within a few picoseconds as the cross-alignment dephases.
Permuting the pulse polarizations or shifting the THz phase by $\pi$ switches its sign together with that of $\chi$.
Since this transient follows the envelopes of the pump pulses, probing it requires no phase locking between excitation and probe, only control of their delay.
Once the cross-alignment has dephased, the persistent time-odd contribution is not accessible to PECD.
It can, however, be mapped back onto a cross-alignment:
A delayed non-resonant optical pulse, linearly polarized along $X$, transiently aligns the molecules such that $\braket{n_c^2-n_b^2}\neq0$, and the net rotation about $a$ turns this achiral alignment into $\braket{n_bn_c}$, cf. the first term of Eq.~\eqref{eq:kinematic}.
The resulting LPECD peaks about \SI{70}{fs} after the pulse maximum, reaching \SI{2.6}{\%} and \SI{0.9}{\%} for the two schemes of Fig.~\ref{fig:nonresonant_pecd}, at the expense of part of $\braket{\hat J_a}$.
Since a single linearly polarized pulse is achiral, this signal changes sign with the handedness of the ensemble.
PECD can also probe the wave packet created by microwave excitation.
There, the cross-alignment is proportional to $\mathrm{Re}\,\chi$, such that the LPECD oscillates with $\omega_\chi$, reaching about \SI{15}{\%}, and vanishes when the chirality is purely time-odd.
Hence, the ionizing pulse needs to be synchronized with the microwave pulses.

\paragraph*{Conclusions}
We have shown how to prepare achiral molecules in rotationally chiral states via angular momentum localization in the molecular frame, without the need for molecular alignment or orientation.
The chirality comprises a time-odd, falsely chiral and a time-even, truly chiral contribution, linked by Euler's rotation equations.
Creating either contribution with net enantiomeric excess requires three non-coplanar polarization directions, analogously to the requirements for enantio-selective excitation in three-wave mixing~\cite{Leibscher2019}.
The handedness of the induced chirality can be controlled by the handedness of the excitation sequence.
Whereas the chirality generated by microwave excitation alternates between time-odd and time-even character, and a phase-sensitive microwave probe resolves the two contributions separately,  only the time-odd contribution persists
after THz and optical excitation.
Similarly to laser-induced vibrational chirality~\cite{Tikhonov2022}, time-delayed ionization and measurement of PECD detects the transient time-even contribution and a delayed non-resonant pulse renders the persistent time-odd contribution accessible to PECD as well.
Indeed, our simulations have shown that purely rotational chiral dynamics gives rise to PECD of similar strength as typically reported for chiral molecules \cite{Sparling2025}.
The suggested pump-probe spectroscopy of rotational chirality can be realized with existing experimental capabilities both in the resonant microwave regime and with combinations of optical pulses and THz radiation.
By establishing rotational chirality in the molecular frame, together with probes that separate its time-odd and time-even contributions, our approach provides a practical framework for investigating time-odd chiroptical phenomena. Since intermolecular interactions depend on the rotational states of the collision partners, controlling rotational handedness could also offer a means to steer enantio-sensitive reactions far from equilibrium.

\section*{Acknowledgements}
We thank the anonymous referee for suggesting to detect the induced rotational chirality with a three-wave mixing scheme, which led to the microwave probe proposed here.
Funded by the Deutsche Forschungsgemeinschaft (DFG, German Research Foundation) – Projektnummer 328961117 – SFB 1319 ELCH.
This research was supported in part by grant NSF PHY-2309135 to the Kavli Institute for Theoretical Physics (KITP).

\section*{Supporting Information}
Selected molecular-frame angular momentum distributions for \ce{COFCl};
derivation of the conditions for laser-induced rotational chirality;
computational details (PDF).

\printbibliography

\end{document}

\maketitle

\begin{figure}[tbh]
\centering
\includegraphics[width=\textwidth]{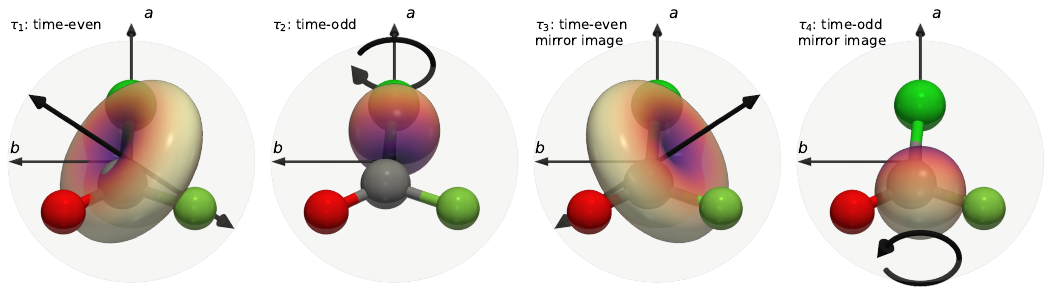}
\caption{
Molecular-frame angular momentum distribution of the rotational wave packet of \ce{COFCl} prepared by the pulse sequence of Fig.~2 of the main text, at four the indicated delays $\tau_1$ to $\tau_4$.
The distribution's color and distance from the molecule's center of mass show how likely the angular momentum is to point along a certain molecular-frame direction, with brighter color and larger radius indicating higher probability.
The gray ellipse marks the molecular plane, spanned by the $a$ and $b$ axes.
At $\tau_1$, the chirality is time-even: the distribution forms a torus around an axis tilted by \SI{45}{\degree} out of the molecular plane (double-headed arrow). Both senses of rotation about this axis are equally probable; the distribution has a definite tilt but no definite sense of rotation.
At $\tau_2$, the chirality is time-odd: the distribution is a single lobe along $+a$, with a definite sense of rotation about $a$ (curled arrow).
At $\tau_3$ and $\tau_4$, the distributions are the mirror images of those at $\tau_1$ and $\tau_2$.
}
\label{fig:S_Jdistribution}
\end{figure}

\section{Conditions for laser-induced rotational chirality}
\label{app:conditions}
Deriving the conditions for generating rotational chirality in an achiral asymmetric rigid rotor proceeds analogously to 
the case of creating a chiral rovibrational wave packet~\cite{Leibscher2024}.
The rotational state is expressed in the basis of asymmetric top eigenstates,
$\ket{j}\equiv\ket{J_{K_a,K_c}, M}$ with $J=0,1,2,\dots$ the rotational quantum number, $M=-J, \dots, J$ the projection of the angular momentum along the laboratory axis, and $K_a=0,1,\dots,J$ and $K_c=0,1,\dots,J$ labeling the symmetric-top basis functions in the prolate and oblate limits~\cite{Bunker2012}.
We use the same axis assignment as for the near prolate molecules discussed in the main text. The oblate case follows by cyclic permutation of $a$,$b$ and $c$.
For the simplest case of a superposition of two states, 
$\ket{j_1}$ and $\ket{j_2}$, the condition for a chiral wave packet becomes
\begin{align}
\braket{\hat J_a(t)} = 
2 \mathrm{Re}\left[
a_{j_1}(t) a^*_{j_2}(t)
\braket{j_2 | \hat J_a | j_1}
\right]
\neq 0
\;.
\label{eq:exp_Ja}
\end{align}
Thus, both the matrix element $\braket{j_2|\hat J_a|j_1}$ and the two expansion coefficients must be nonzero. The former determines which rotational states can form a chiral superposition, while the latter determines how these states can be populated.

The states $\ket{j_1}$ and $\ket{j_2}$ transform according to the irreducible representations of the asymmetric top symmetry group $D_2$. A nonzero matrix element $\braket{j_2|\hat J_a|j_1}$ requires the direct product of the corresponding representations to be totally symmetric,
\begin{align*}
\Gamma(\ket{j_2}) \times \Gamma(\hat J_a) \times \Gamma(\ket{j_1})
= A
\;.
\end{align*}
$\hat J_a$ transforms as $\Gamma(\hat J_a)=B_a$, while the symmetry of the rotational states is determined by the parities of $K_a$ and $K_c$, as summarized in Table~\ref{tab:D2_character_table}.
Using $B_\alpha\times B_\alpha=A$, $A\times B_\alpha=B_\alpha$, as well as $B_a\times B_b= B_c$ and cyclic permutations~\cite{Bunker2012}, non-zero $\braket{j_2 | \hat J_a | j_1}$ requires superposition of states with either $A$ and $B_a$ symmetry or $B_b$ and $B_c$ symmetry.
The time-even pseudoscalar $\{\hat J_b,\hat J_c\}$ transforms as $B_b\times B_c=B_a$, i.e., like $\hat J_a$, so that the same superpositions carry both contributions to the chirality, Eq.~(2) in the main text.
In addition, orthogonality of the asymmetric-top eigenstates requires the two states to have the same $J$ and $M$.

\begin{table}
\centering
\caption{
Character table for the rotational symmetry group $D_2$ and symmetry properties of the asymmetric-top eigenstates and principal-axis components of the angular momentum operator \cite{Bunker2012}. Here, $\hat R_i^\pi$ denotes a rotation by $\pi$ about the $i$th principal axis; e and o denote even and odd parity of $K_a$ and $K_c$, respectively.
}
\label{tab:D2_character_table}
\vspace*{1ex}
\begin{tabular}{lrrrrccr}
\hline
&\;\;\;\; $\hat{E}$ &\;\;\;\; $\hat{R}_a^\pi$ &\;\;\;\; $\hat{R}_b^\pi$ &\;\;\;\; $\hat {R}_c^\pi$ &\;\;&\;\; $K_aK_c$  \;\;& $\hat J_i$\\
\hline
$A$ & 1 & 1 & 1 & 1 & & ee  &\\
$B_a$ & 1 & 1 & -1 & -1 & & eo & $\hat J_a$\\
$B_b$ & 1 & -1 & 1 & -1 & & oo & $\hat J_b$\\
$B_c$ & 1 & -1 & -1 & 1 & & oe & $\hat J_c$\\
\hline
\end{tabular}
\end{table}

Next we determine the number and polarization directions of the fields required to populate a chiral superposition.
These are constrained by symmetry of the rotational states with respect to both the molecule-fixed and laboratory-fixed axes.
Starting from an initial state $\ket{j_0}$, time-dependent perturbation theory gives a nonzero amplitude for a state $\ket{j}$ if at least one sequence of $l$ interactions satisfies
\begin{eqnarray*}
\braket{j|\hat H_{\mathrm{int}}^{(l)}|j_0}
&\equiv&
\braket{j|\hat H_{\mathrm{int}}|j^{(l-1)}}
\cdots
\braket{j^{(2)}|\hat H_{\mathrm{int}}|j^{(1)}}
\braket{j^{(1)}|\hat H_{\mathrm{int}}|j_0}
\neq0,
\end{eqnarray*}
where $\ket{j^{(i)}}$ are intermediate rotational states and $\hat H_{\mathrm{int}}$ refers to the electric dipole interaction. For the two states in Eq.~\eqref{eq:exp_Ja}, such pathways must exist for some orders $l_1$ and $l_2$. In terms of the $D_2$ irreducible representations, this condition reads ($i=1,2$)
\begin{align*}
\Gamma(\ket{j_i})
\times
\Gamma(\hat H_{\mathrm{int}}^{(l_i)})
\times
\Gamma(\ket{j_0})
&=A \;,
\end{align*}
and combination with the symmetry requirements on $\ket{j_1}$ and $\ket{j_2}$
yields
\begin{align}
\Gamma(\hat H_{\mathrm{int}}^{(l_1)})
=
B_a\times
\Gamma(\hat H_{\mathrm{int}}^{(l_2)}) \;.
\label{eq:conditions_for_fields}
\end{align}
Thus, the symmetries of the excitation pathways must be either $A$ and $B_a$, or $B_b$ and $B_c$.

The dipole interaction components transform as
$\Gamma(\hat H_{\mathrm{int},\alpha})=B_\alpha$ (with $\alpha=a,b,c)$ since
$\hat H_{\mathrm{int},\alpha}\propto \hat d_\alpha$~\cite{Leibscher2024}. For achiral molecules, which in the prolate case have $\hat d_c=0$, Eq.~\eqref{eq:conditions_for_fields} therefore requires at least three dipole interactions,  $l_1 + l_2 \geq 3$, cf. $l_1=1$ and $l_2=2$ for the example of creating rotational chirality with three microwave pulses.
Notably, purely non-resonant excitation, where the interaction is mediated by the polarizability tensor, does not provide the required symmetry: For achiral molecules, the non-vanishing irreducible polarizability tensor elements transform as $A$ or $B_c$. These components cannot satisfy Eq.~\eqref{eq:conditions_for_fields} on their own; non-resonant excitation must therefore be combined with at least one dipole interaction, which can be realized e.g. by a THz pulse or a static electric field.
Moreover, the laboratory frame symmetry of rotational states imposes a constraint on the polarization directions. For $Z$-polarized fields, rotational transitions obey $\Delta M=0$, whereas for $X$- and $Y$-polarized fields $\Delta M=\pm1$. For an initial state with quantum numbers $J_0$ and $M_0$, consider, without loss of generality, a first-order transition driven by $Z$-polarized light, which produces a state with $J=J_0+1$ and $M=M_0$. A second-order pathway to the same $J$ and $M$ requires at least one $X$- and one $Y$-polarized interaction \cite{Leibscher2019}. Consequently, the minimal excitation sequence for exciting a chiral superposition requires polarization components along all three laboratory axes, i.e., three non-coplanar polarization directions.

\section{Computational details}

\subsection{Rotational dynamics}
For the simulation of rotational dynamics under microwave excitation, we have solved the time-dependent Schrödinger equation for a rigid asymmetric rotor in the electric dipole approximation, initially in its ground state, using the QDYN library \cite{qdyn}.
The excitation with non-resonant and THz pulses is instead modeled classically, employing a Monte-Carlo simulation as described in e.g. Ref.~\cite{Tutunnikov2018}, to solve Euler's equation for an ensemble of $N=10^6$ classical rotors, initially randomly oriented and with a thermal distribution of angular frequencies, corresponding to a temperature of \SI{5}{K}.
For simulation times shorter than the revival time (\SI{14}{ps} for \ce{CH2O}), the results of quantum mechanical and classical calculations are expected to coincide~\cite{Tutunnikov2021}.
THz pulses are modeled as $E(t)=E_0 (1-2at^2) e^{-at^2}$ with amplitude $E_0=\SI{8}{MV/cm}$ and $a=\SI{3.06}{ps^{-2}}$ \cite{Tutunnikov2021}.
The optical pulses are taken as Gaussian envelopes with \SI{2e+8}{V/cm} peak amplitude and \SI{0.2}{ps} full width at half maximum of the intensity profile. The optical pulses have been time-averaged over an optical cycle \cite{Tutunnikov2018}.
The delayed optical pulse in Fig.~3 of the main text has the same parameters as the optical pump pulses.
Rotational constants, permanent dipole moments and polarizability components for \ce{CH2O} and \ce{COFCl} have been calculated with MOLPRO \cite{werner2012molprowires,werner2020molpro} at the equilibrium geometries optimized on the HF/aug-cc-pVTZ level of theory \cite{augccpvdz,Woon1993}.

\subsection{Photoionization}
To calculate the PECD for the induced rotational chirality, we follow the approach of Ref.~\cite{Tikhonov2022}. Briefly, one-photon ionization by a femtosecond extreme-ultraviolet probe pulse is modeled by first-order time-dependent perturbation theory in the frozen-core static-exchange approximation \cite{Gianturco1994,Natalense1999}.
The corresponding ionization amplitude for a molecule with orientation $\omega$ and photoelectron momentum $\vec k$ is denoted by $a_{\vec k}(t,\omega)$.
The transition dipole matrix elements into the photoelectron continuum have been calculated with ePolyScat \cite{Gianturco1994,Natalense1999} from the optimized equilibrium geometries of \ce{COFCl} and \ce{CH2O}. Neglecting centrifugal distortion is appropriate for the rotational energies considered here, for which the deformation of the nuclear scaffold is negligible.

We assume the probe pulse to be short compared to the timescale of the rotational dynamics. The temporal shape of the probe can then be reasonably approximated by a delta pulse and the rotational wave function, $\psi(t,\omega)$, can be regarded as frozen during the ionization process.
Consequently, the orientation averaged ionization amplitude takes the form of a classical orientation average, Eq.~(4) of the main text, with orientational probability distribution $P_\mathrm{rot}=|\psi(t,\omega)|^2$.
Despite the classical appearance of the orientation average, $\psi(t,\omega)$ contains the coherent superposition of rotational states and hence the time-dependent interference responsible for the rotational chirality.
When expanded in rotational eigenstates, $\psi(t,\omega)=\sum_j c_j(t)\psi_j(\omega)$, Eq.~(4) of the main text becomes
\begin{eqnarray}
\label{eq:ionization_amplitude_expanded}
a_{\vec k}(t)
&=&
\sum_j
|c_j|^2
\int a_{\vec k}(t,\omega)
|\psi_j(\omega)|^2 \mathrm{d}\omega
\nonumber\\
&&+
2\sum_{i>j}
|c_i c_j|
\int a_{\vec k}(t,\omega)
\mathrm{Re}
\left[e^{i(\omega_{ij} t + \Phi_{ij})} \psi_i(\omega)^* \psi_j(\omega)\right]
\mathrm{d}\omega \;,
\end{eqnarray}
where we have used that, after the pump, $c_j(t)=|c_j|e^{-i(E_jt/\hbar-\varphi_j)}$ with $E_j$ the rotational energies. Here, $\omega_{ij}=(E_i-E_j)/\hbar$ is the transition frequency between the rotational levels $\psi_i(\omega)$ and $\psi_j(\omega)$ and $\Phi_{ij}=\varphi_j-\varphi_i$ represents the phase imprinted onto the rotational states by the pump sequence.
The first term corresponds to the contributions from individual eigenstates. Since these are achiral, it does not contribute to PECD.
Since chirality arises from the mixing terms, only the second term in Eq.~\eqref{eq:ionization_amplitude_expanded} gives rise to PECD.
Its chiral part is, to leading order, the cross-alignment $\braket{n_xn_y}$ of the main text: For the two-state wave packet of Fig.~2 of the main text, the LPECD is exactly proportional to it; for the thermal ensembles of Fig.~3 of the main text, it accounts for more than \SI{98}{\%} of the signal.
Similarly, thermal population of rotational states gives rise to incoherent, isotropic contributions to the photoelectron spectrum, not contributing to PECD.
Thermal population will, however, reduce the degree of the induced chirality and hence the contrast of the PECD signal.

We expect the same mechanism to apply to multi-photon ionization provided that the probe pulse remains short compared with the rotational timescale. In this regime, the probe still samples the instantaneous rotational state, while the details of the electronic ionization amplitude and the associated scattering phase modify the magnitude and phase of the PECD.

%%%%%%%%%%%%%%%%%%%%%%%%%%%%%%%%%%%%%%%%%%%%%%%%%%%%%%%%%%%%%%%%%%%%%
%% If you are using classical BibTeX rather than biblatex,
%% remove the \printbibliography and uncomment the \bibliograpy one
%%%%%%%%%%%%%%%%%%%%%%%%%%%%%%%%%%%%%%%%%%%%%%%%%%%%%%%%%%%%%%%%%%%%%
\printbibliography
%\bibliography{references.bib}